\documentclass{mem}
\usepackage{natbib}
\usepackage{txfonts}
\usepackage{balance}
\usepackage{graphicx}
\usepackage[a4paper]{hyperref}
\idline{1}{1}
\begin{document}
\def\teff{$T\rm_{eff }$}
\def\kms{$\mathrm {km s}^{-1}$}

\title{
Exploring the Physics of Type Ia Supernovae Through the X-ray Spectra of their Remnants
}

    \subtitle{}

\author{
C. \,Badenes\inst{1} 
\and K. \,Borkowski\inst{2}
\and E. \,Bravo\inst{3,4}
\and J.P. \, Hughes\inst{1}
\and U. \,Hwang\inst{5}
}

\offprints{C. Badenes}
 
\institute{
Rutgers University, Department of Physics and Astronomy,
136 Frelinghuysen Rd, Piscataway NJ 08854-8019, USA,
\email{badenes@physics.rutgers.edu}
\and
North Carolina State University, Department of Physics,
Box 8202, Raleigh NC 27965-8202, USA
\and
Departament de F\'isica i Enginyeria Nuclear, 
Universitat Polit\`ecnica de Catalunya,
Diagonal 647, Barcelona 08028, Spain
\and
Institut d'Estudis Espacials de Catalunya, 
Campus UAB, Facultat de Ci\`{e}ncies. Torre C5. 
Bellaterra, Barcelona 08193, Spain
\and
NASA - Goddard Space Flight Center,
Greenbelt, MD 20771, USA
}

\authorrunning{Badenes et al.}

\titlerunning{Type Ia SNRs}

\abstract{
We present the results of an ongoing project to use the X-ray observations of Type Ia Supernova Remnants
to constrain the physical processes involved in Type Ia Supernova explosions. We use the Tycho
Supernova Remnant (SN 1572) as a benchmark case, comparing its observed spectrum with models for
the X-ray emission from the shocked ejecta generated from different kinds of Type Ia explosions.
Both the integrated spectrum of Tycho and the spatial distribution of the Fe and Si emission in 
the remnant are well reproduced by delayed detonation models with stratified ejecta. All the other
Type Ia explosion models fail, including well-mixed deflagrations calculated in three dimensions.
}

\maketitle{}

\section{Introduction}
Despite the considerable efforts of the last decades, the physical mechanism responsible for the
explosion of Type Ia Supernovae (SNe) still remains obscure \citep{hillebrandt00:Ia-review}.
Traditionally, theoretical calculations of Type Ia SN explosions have been constrained mainly
through comparison with optical SN spectra, but the complexity of the radiative transfer calculations
involved has only led to a limited reduction of the multitude of possible explosion models
\citep[see, for instance][]{baron03:detectability}. In \citet{badenes03:xray} and \citet{badenes05:xray}
(henceforth Paper I and Paper II), 
we introduced a new way to explore the physics of Type Ia SN explosions. We have proposed to compare the X-ray
emission from the shocked ejecta in Type Ia Supernova Remnants (SNRs) with synthetic X-ray spectra
generated from hydrodynamic calculations of the interaction of a grid of Type Ia SN explosion models with 
the surrounding ambient medium (AM), coupled to nonequilibrium ionization (NEI) calculations and an X-ray 
spectral code. 

This comparison is possible because, for the first time in history, X-ray observations of  
SNRs with high spatial and spectral
resolution have become available thanks to {\it Chandra} and {\it XMM-Newton}.
Among the SNRs that can be observed with sufficient detail, many are still young, and their X-ray spectra are 
dominated by the emission from the shocked SN ejecta. Several of them are thought to be of Type Ia origin, 
like the Tycho SNR (SN 1572), SN 1006 and the Kepler SNR (SN 1604) in our Galaxy, or 0509-67.5, DEM L71 
and N103B in the Magellanic Clouds. Here, we present the results from comparison between our
models and the X-ray spectrum the Tycho SNR.

\section{Modeling the X-ray Spectrum of the Tycho SNR}

Tycho is one of the brightest X-ray SNRs in the sky, and has been observed in great detail at all
wavelengths from the radio to $\gamma$-rays. In the X-rays, it has been a primary target for all the major 
satellite missions, including {\it ASCA}, {\it Chandra}, and {\it XMM-Newton}. Among the conclusions drawn 
from these observations, we highlight the following: 
\begin{itemize}
\item The outer edge of emission, which coincides in the X-rays and in radio observations, can be identified with 
the forward shock (FS).
\item The integrated X-ray spectrum of the SNR is dominated by strong emission lines from Fe, Si, S, Ar, Ca and 
other elements.
\item The X-ray emission from the shocked AM just behind the FS is a featureless continuum 
\citep{hwang02:tycho-rim,warren05:Tycho}. This implies that all the X-ray line emission has to come 
from the shocked ejecta.
\item The peak of the Fe K$\alpha$ line emission is at lower radii than that of the Fe L or Si He$\alpha$ emission
\citep{hwang97:tycho-ASCA}. This Fe K$\alpha$ emission comes from material with a higher temperature and a lower
ionization timescale than the rest of the ejecta \citep{hwang98:tycho-Feemission}.
\end{itemize}

\begin{figure}
  \centering
  \includegraphics[clip=true,scale=0.6]{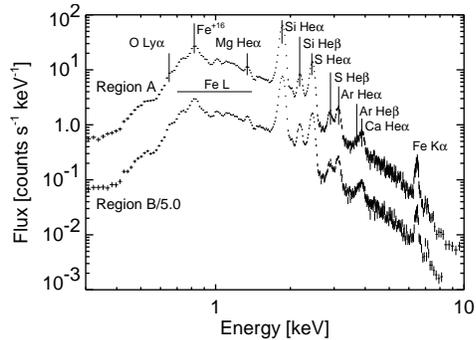}
  \caption{\footnotesize
    Spatially integrated {\it XMM-Newton} EPIC-MOS spectra extracted from two regions (A and B) in the western side of the 
    Tycho SNR. Region A encompasses approximately 40\% of the SNR, and region B is a subset of region A. The most 
    important emission lines have been labeled for clarity.
  }
  \label{fig-1}
\end{figure}

The {\it XMM-Newton} EPIC MOS spectrum of the Tycho SNR is presented in Figure \ref{fig-1}. As shown in 
\citet{warren05:Tycho}, the continuum emission from the AM is produced by a FS whose dynamics is strongly
modified by cosmic ray (CR) acceleration \citep[see][]{decourchelle00:cr-thermalxray}. 
This brings the contact discontinuity between shocked ejecta and shocked AM very close to the FS. The reverse
shock, on the other hand, is deep into the ejecta, and the material close to it is very hot (as shown by the
morphology of Fe K$\alpha$ emission), so CR acceleration at the reverse shock seems unlikely.

\begin{figure*}
  \includegraphics*[clip=true,scale=0.5]{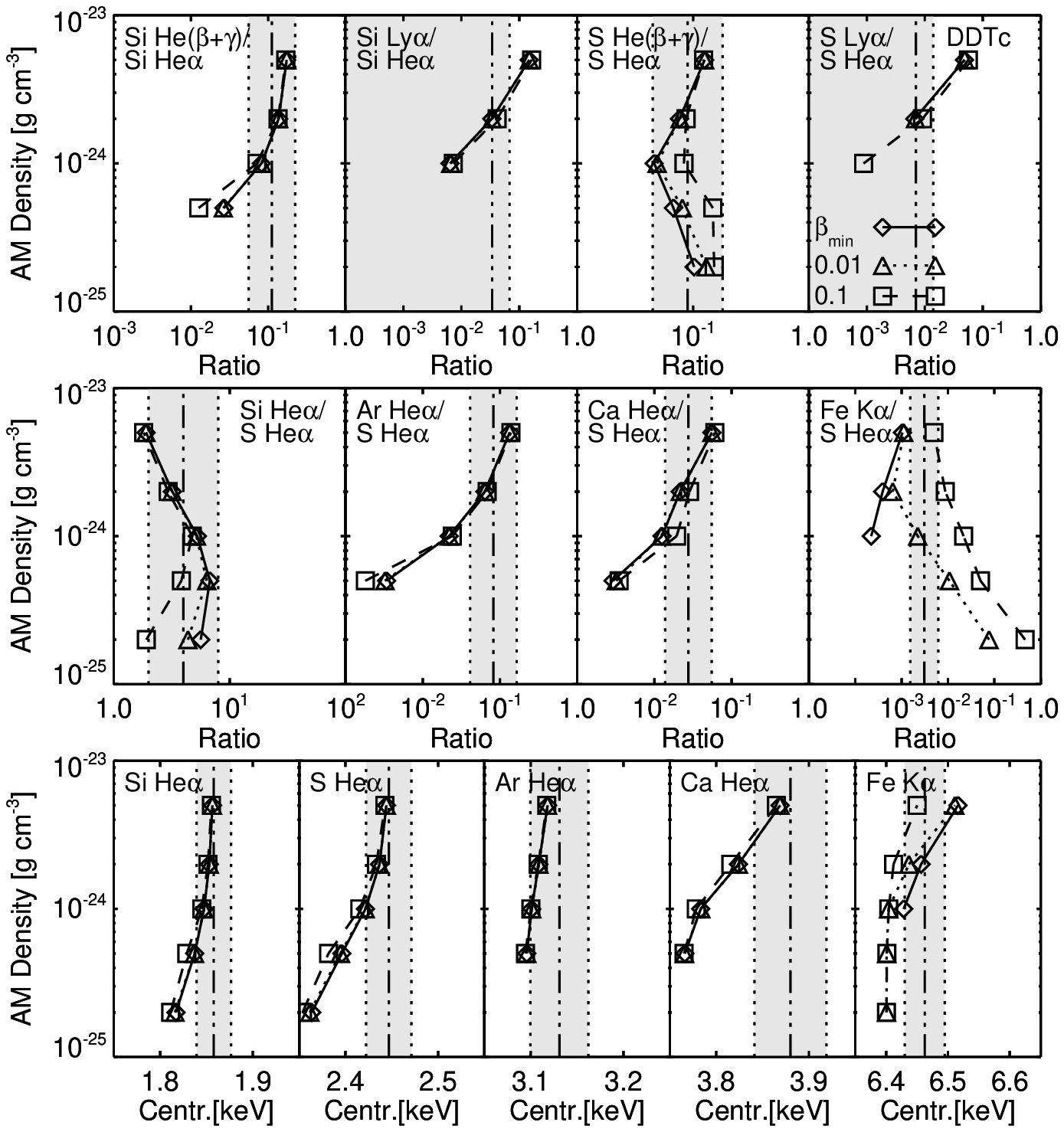}
  \includegraphics*[clip=true,scale=0.5]{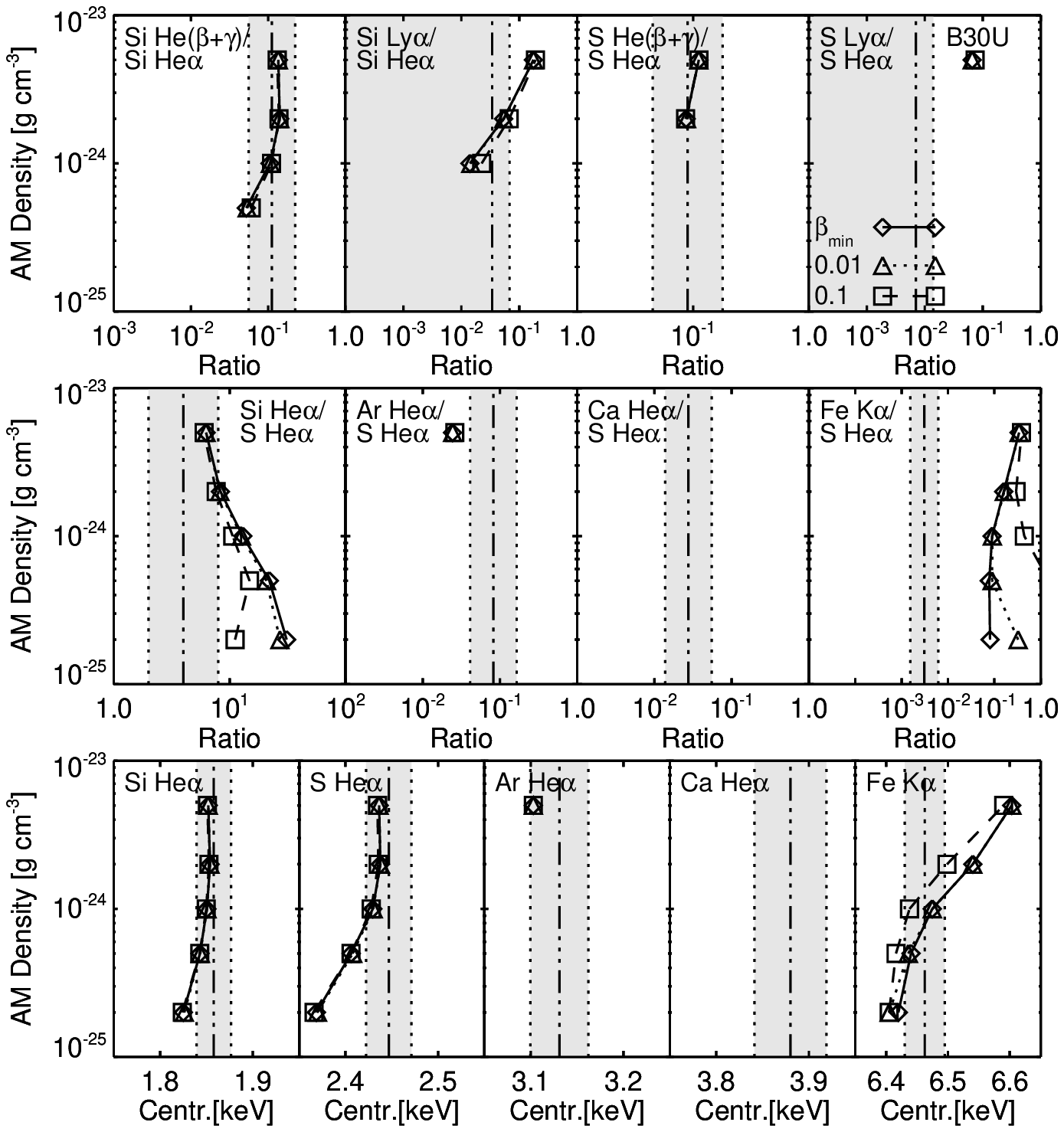}
  \caption{\footnotesize
    Line flux ratios and line centroids predicted for the delayed detonation model DDTc (left) and the
    three-dimensional deflagration model B30U (right). Values are given as a function of 
    $\rho_{AM}$ (vertical axes), and for three different values of $\beta$: $\beta_{min}$ (diamonds, solid line), 0.01 
    (triangles, dotted line), and 0.1 (squares, dashed line). If a particular flux ratio or centroid is missing, 
    this indicates that the relevant line in the synthetic spectrum is either very weak or altogether absent. The 
    values determined from the {\it XMM-Newton} spectrum of the Tycho SNR are 
    plotted as vertical lines, and tolerance ranges are given, with allowed regions shaded in grey.
  }
  \label{fig-2}
\end{figure*}

Using these observational results as a starting point, we attempt to model the X-ray emission from the shocked ejecta
with the models described in Papers I and II. Since all the X-ray line emission comes from the shocked ejecta, 
we have used the line flux ratios and line centroids determined from the {\it XMM-Newton} spectrum to reduce the 
dimensionality of the problem and identify the most promising models. This is illustrated in Figure \ref{fig-2}, 
where we compare the line emission 
from the Tycho SNR with the synthetic spectra for the shocked ejecta at a SNR age of $t=430$ yr of two Type Ia SN 
models, a one-dimensional delayed detonation (DDTc) and a deflagration calculated in three dimensions (B30U). 
Each panel of Fig. \ref{fig-2} corresponds to one of the eight line flux ratios (top two rows of panels) and five line 
centroids (bottom row of panels) that we have considered, for a total of thirteen observable parameters. 
The observed values are represented by vertical lines, with appropriate tolerance ranges and allowed regions 
shaded in grey. For each Type Ia explosion model, we vary the values of $\rho_{AM}$ (the density of the AM that 
is interacting with the ejecta) and $\beta$ (the amount of collisionless heating of electrons at the reverse shock,
defined as the ratio of electron to ion postshock specific internal energy $\varepsilon_{e,s}/\varepsilon_{i,s}$) 
within ranges that are reasonable for the Tycho SNR. See Papers I and II for details on the Type Ia models, 
definitions of the parameters and a complete explanation of the simulation scheme. 

Models DDTc and B30U are provided here just as representative examples - a more detailed exploration of the 
parameter space for Type Ia SN explosions will be given in the forthcoming refereed paper. As can be seen in
Fig. \ref{fig-2}, model DDTc is much more successful than B30U. At 
$\rho_{AM}=2 \cdot 10^{-24} \, \mathrm{g \cdot cm^{-3}}$ and $0.01 < \beta < 0.1$, model DDTc is
capable of reproducing twelve of the thirteen observable parameters, the only exception being the centroid of the 
Ca He$\alpha$ line. Model B30U, on the other hand, cannot reproduce even the fundamental properties of the 
Fe and Si line emission for any combination of $\rho_{AM}$ and
$\beta$, with important lines like Ar He$\alpha$ and Ca He$\alpha$ being altogether absent from the 
synthetic spectra. When this kind of comparison is performed using other models, the conclusion is that
only one-dimensional delayed detonations have any hope of reproducing the fundamental characteristics
of the line emission from the Tycho SNR. All other models fail, including one-dimensional deflagrations, 
sub-Chandrasekhar explosions and pulsating delayed detonations. The three-dimensional deflagration models
with well-mixed ejecta also fail, because the absence of stratification in the ejecta cannot account for the
different spectral properties of Fe and Si (see discussion in Section 3 of Paper II).

\begin{figure*}
  \includegraphics*[clip=true,scale=1.05]{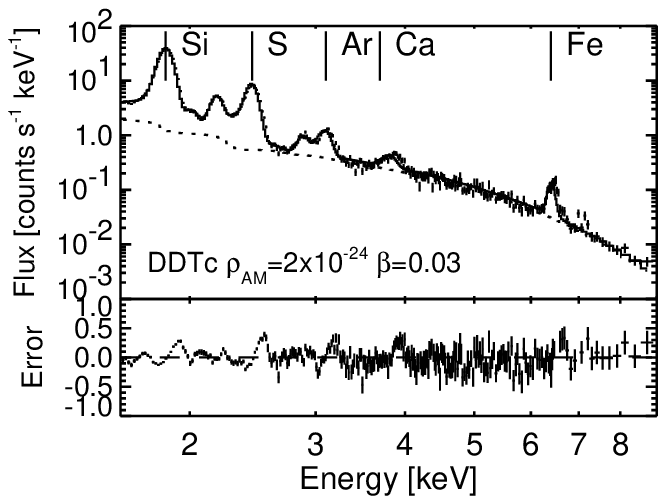}
  \includegraphics*[clip=true,scale=1.05]{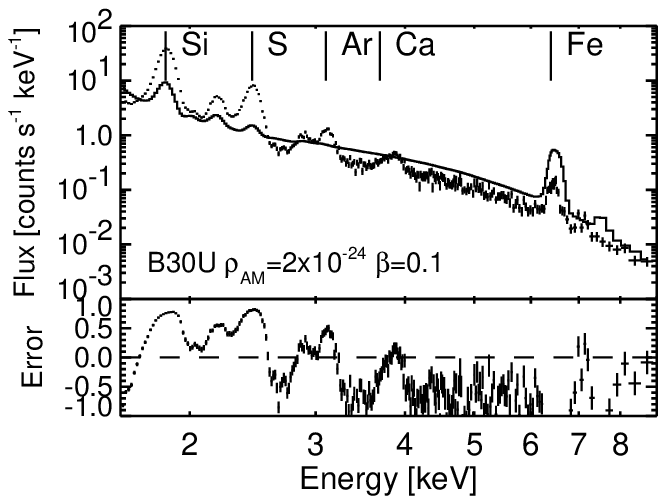}
  \caption{\footnotesize
    Comparison of the synthetic spectra with the best combination of $\rho_{AM}$ (in $\mathrm{g \cdot cm^{-3}}$) 
    and $\beta$, for models DDTc (left) and B30U (right) with the {\it XMM-Newton} observation of the Tycho SNR.
    The dotted line in the DDTc plot represents the power law continuum with index $2.72$ 
    \citep[after][]{fink94:tycho-ginga} added to model the contribution from the shocked AM emission. The continuum from the
    shocked ejecta in model B30U is so high that the fit did not admit an additional power law component.
  }
  \label{fig-3}
\end{figure*}

Once the most promising values of $\rho_{AM}$ and $\beta$ have been identified through the properties of the line emission, 
the entire synthetic spectra can be compared with the observations to evaluate the performance of the models. In Figure
\ref{fig-3}, we plot the synthetic spectra of model DDTc (with $\rho_{AM} = 2 \cdot 10^{-24} \, \mathrm{g \cdot cm^{-3}}$
and $\beta=0.03$) and model B30U (with $\rho_{AM} = 2 \cdot 10^{-24} \, \mathrm{g \cdot cm^{-3}}$
and $\beta=0.1$), alongside the integrated {\it XMM-Newton} spectrum. The fit obtained with model DDTc is quite 
impressive, with absolute errors comparable to the quality of the atomic data in the X-ray spectral codes.
It is worth noting that the normalization of the flux in the synthetic spectra to the observed flux yields an 
estimate for the distance $D$ to the SNR. This 'normalization distance' $D_{norm}$ is important, because the
effect of CR acceleration at the FS makes it very difficult to constrain $D$ by methods that rely on
the dynamics of the FS. In the case of the DDTc model presented here, $D_{norm} = 2.7$ kpc, a value that is
within the $1.5 < D < 3$ kpc range proposed by \citet{smith91:six_balmer_snrs}. 

\begin{figure}
  \centering
  \includegraphics[clip=true,scale=0.5]{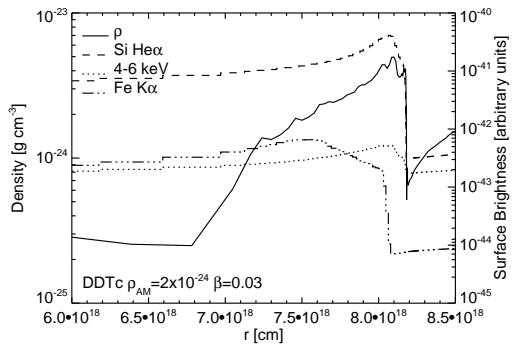}
  \caption{\footnotesize
    Surface brightness profiles of the Si He$\alpha$ line, 4 to 6 keV continuum and the Fe K$\alpha$ line
    for model DDTc with $\rho_{AM} = 2 \cdot 10^{-24} \, \mathrm{g \cdot cm^{-3}}$
    and $\beta=0.03$. The region of the model represented here corresponds mostly to the shocked ejecta:
    the reverse shock is at a radius of $\sim 7.2 \cdot 10^{18}$ cm and the contact discontinuity at 
    $\sim 8.2 \cdot 10^{18}$ cm.
  }
  \label{fig-4}
\end{figure}

Any successful model for the X-ray emission from the Tycho SNR must be able to reproduce the spatial
distribution of the line emission as well as the integrated spectrum. In particular, it has to provide an explanation 
for the fact that the Fe K$\alpha$ line peaks interior to both Fe L and Si He$\alpha$. 
In Section 2.4 of Paper II we proposed that collisionless electron heating at the reverse shock could provide such 
an explanation, by increasing the electron temperature in the
innermost shocked ejecta and exciting Fe K$\alpha$ emission close to the reverse shock. We illustrate this  
in Figure \ref{fig-4}, where we plot the surface brightness profile of the Si He$\alpha$ and Fe K$\alpha$ lines, along
with that of the 4 to 6 keV continuum emission for model DDTc with $\rho_{AM} = 2 \cdot 10^{-24} \, \mathrm{g \cdot cm^{-3}}$
and $\beta=0.03$. Qualitatively, these surface brightness profiles are similar to those obtained by {\it XMM-Newton}
\citep{decourchelle01:tycho-xmm} and {\it Chandra} \citep{warren05:Tycho}, but a detailed quantitative comparison with the
data is required in order to draw conclusions. This issue will also be addressed in the forthcoming paper.

We stress that, even though our analysis of the ejecta emission is based on adiabatic one-dimensional models,
the approximations we have made are well justified. \citet{sorokina04:typeIasnrs} have claimed
that thermal conduction and radiative losses in the SN ejecta need to be taken into account for Type Ia SNRs, but 
there is no observational evidence to support these claims. No trace of radiatively cooled ejecta has been found Tycho 
\citep[see][]{smith91:six_balmer_snrs}, and the morphology of the Fe K$\alpha$ emission can only be explained if 
there is a temperature gradient in the shocked ejecta, which is incompatible with efficient thermal conduction.

\section{Conclusions}

In this contribution, we have summarized the main results of our ongoing efforts to use the X-ray emission
from the Tycho SNR to constrain the physics of Type Ia SN explosions. 
These results are very promising: we have found that only one class of models, one-dimensional 
delayed detonations, is capable of reproducing the fundamental properties of the ejecta emission.
Among the delayed detonations, the preferred model is DDTc, with $E_{k}=1.16 \cdot 10^{51}$ erg
and 0.8 solar masses of $^{56}$Ni. This model provides a good approximation to the 
integrated X-ray spectrum for $\rho_{AM} = 2 \cdot 10^{-24} \, \mathrm{g \cdot cm^{-3}}$ and $\beta=0.03$. Other
constraints, like the surface brightness profile of the most relevant lines, the normalization distance $D_{norm}$, 
and the apparent radii of the contact discontinuity
and the reverse shock, are also in agreement with the observations \citep[see][]{warren05:Tycho}. More
details will be provided in the forthcoming refereed paper.

It is remarkable that the sophisticated three-dimensional deflagration models with well-mixed ejecta published 
by several groups \citep[see][for a review]{bravo04:3D_review} fail to reproduce the fundamental properties of the X-ray 
emission from Tycho, in particular the fact that Fe and Si have different spectral properties. This indicates that 
some degree of elemental stratification must exist in Type Ia SN ejecta. The ultimate physical cause of this 
stratification remains an open issue, and even though some explanations have been proposed, no self-consistent
model for Type Ia explosions exists yet. 

\begin{acknowledgements}
I was able to attend this conference funded by a {\it Chandra} GO grant from the Smithsonian
Astrophysical Observatory (PI: J.Hughes). 
\end{acknowledgements}

%Bibliography-related commands
%\bibliographystyle{aa}
%\bibliography{/home/carles/docs/mybibliography}

\end{document}